\documentclass[journal=jacsat,manuscript=article]{achemso}
\usepackage[version=3]{mhchem} 
\usepackage{graphicx}
\usepackage{dcolumn}
\usepackage{bm}
\usepackage{esvect}
\usepackage{amsbsy}
\usepackage{amssymb}
\usepackage{amsmath}
\usepackage[makeroom]{cancel}
\usepackage{graphicx}
\usepackage{grffile}
\graphicspath{ {./} }
\usepackage{caption}
\usepackage{subcaption}
\usepackage{float}
\usepackage{diffcoeff}
\usepackage{physics}
\usepackage[colorinlistoftodos]{todonotes}
\usepackage[colorlinks=true, allcolors=blue]{hyperref}
\usepackage{chemformula}
\usepackage{indentfirst}
\usepackage{color}
\usepackage{comment}
\usepackage{mathrsfs}
\usepackage{slashed}
\usepackage{hyphenat}
\usepackage{natbib}
\usepackage{multirow}
\usepackage{booktabs}
\usepackage{threeparttable}
\author{Shuo Tao}
\affiliation{Department of Physics, Rutgers University, Newark, NJ 07102, United States of America}
\author{Li Zhu}
\affiliation{Department of Physics, Rutgers University, Newark, NJ 07102, United States of America}
\email{li.zhu@rutgers.edu}

\title{EOSnet: Embedded Overlap Structures for Graph Neural Networks in Predicting Material Properties}

\abbreviations{ML,HDNN,ANN,GNN,MLP,FCN}

\begin{document}






\begin{abstract}
\begin{figure}
    \centering
\includegraphics[width=2in]{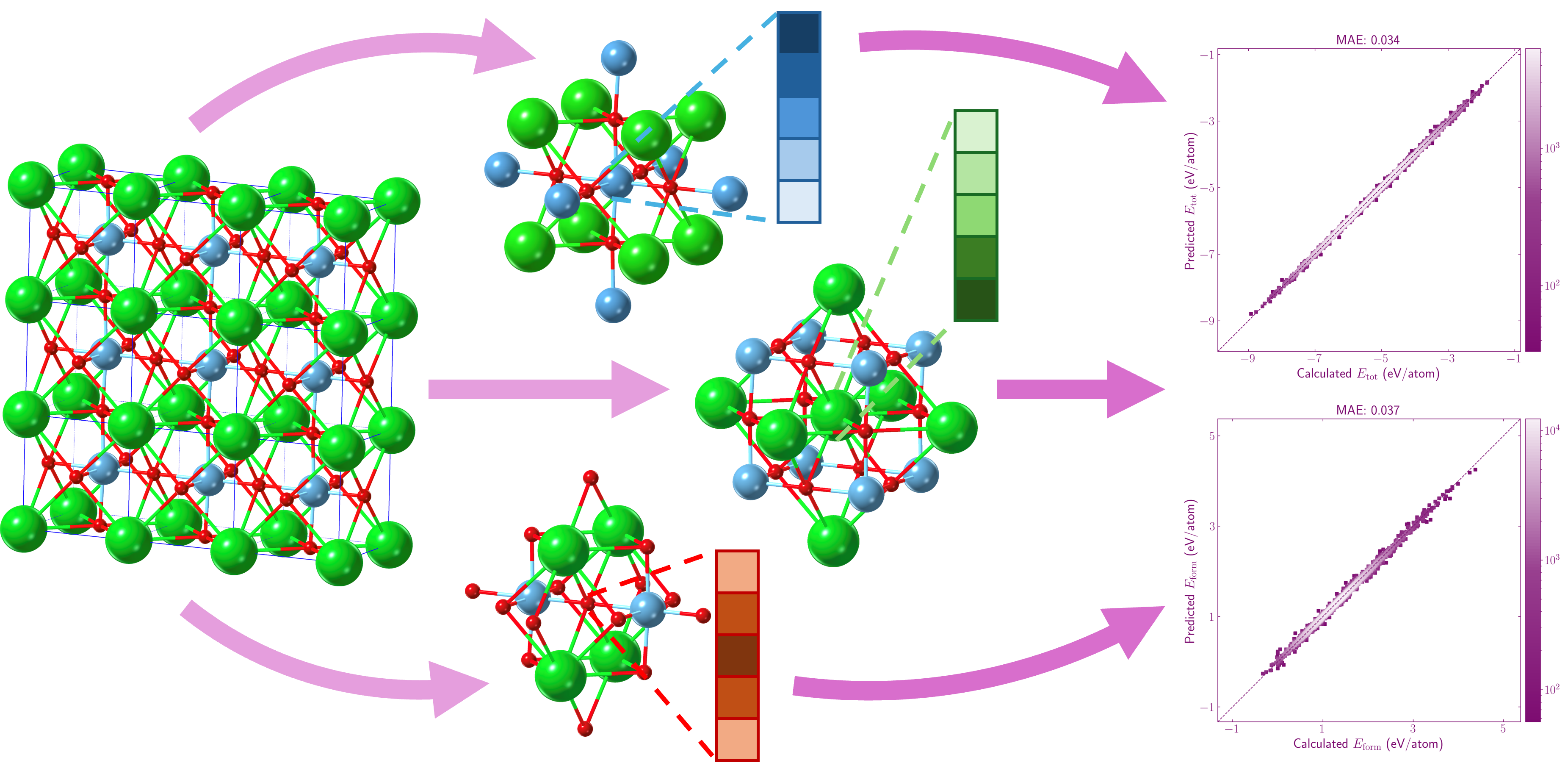}
\end{figure}
Graph Neural Networks (GNNs) have emerged as powerful tools for predicting material properties, yet they often struggle to capture many-body interactions and require extensive manual feature engineering. Here, we present EOSnet (Embedded Overlap Structures for Graph Neural Networks), a novel approach that addresses these limitations by incorporating Gaussian Overlap Matrix (GOM) fingerprints as node features within the GNN architecture. Unlike models that rely on explicit angular terms or human-engineered features, EOSnet efficiently encodes many-body interactions through orbital overlap matrices, providing a rotationally invariant and transferable representation of atomic environments. The model demonstrates superior performance across various materials property prediction tasks, achieving particularly notable results in properties sensitive to many-body interactions. For band gap prediction, EOSnet achieves a mean absolute error of 0.163 eV, surpassing previous state-of-the-art models. The model also excels in predicting mechanical properties and classifying materials, with 97.7\% accuracy in metal/non-metal classification. These results demonstrate that embedding GOM fingerprints into node features enhances the ability of GNNs to capture complex atomic interactions, making EOSnet a powerful tool for materials discovery and property prediction.
\end{abstract}


Machine learning (ML) has become an indispensable tool in the fields of materials science and condensed matter physics, enabling the efficient prediction and discovery of material properties that would otherwise be time-consuming using first-principles calculations. As the demand for high-throughput materials screening has grown, ML models, particularly deep learning architectures \cite{artrithImplementationArtificialNeuralnetwork2016,unkePhysNetNeuralNetwork2019}, have evolved to handle increasingly complex datasets, capturing intricate relationships between atomic structures and their properties. Within this context, Graph Neural Networks (GNNs) \cite{wuComprehensiveSurveyGraph2021,reiserGraphNeuralNetworks2022} have emerged as particularly well-suited for modeling atomic and molecular systems due to their ability to process graph-based data, where atoms and their interactions are represented as nodes and edges, respectively.

In the early applications of ML to materials science, models such as high-dimensional neural networks (HDNNs) \cite{behlerFourGenerationsHighDimensional2021} were employed to learn the complex potential energy surface (PES) across large atomic systems. These models utilized various descriptors like Atom-Centered Symmetry Functions (ACSFs)\cite{behlerGeneralizedNeuralNetworkRepresentation2007,gasteggerWACSFWeightedAtomcentered2018}, $\mathrm{SO}(3)$ power spectrum or Smooth Overlap of Atomic Positions (SOAP) \cite{bartokGaussianApproximationPotentials2010,bartokRepresentingChemicalEnvironments2013}, $\mathrm{SO}(4)$ bispectrum or Spectral Neighbor Analysis Potential (SNAP) \cite{thompsonSpectralNeighborAnalysis2015}, and Atomic Cluster Expansion (ACE)\cite{drautzAtomicClusterExpansion2019,lysogorskiyPerformantImplementationAtomic2021,bochkarevEfficientParametrizationAtomic2022,lysogorskiyActiveLearningStrategies2023}. Each descriptor had its own trade-offs. For example, ACSFs required parameter tuning \cite{imbalzanoAutomaticSelectionAtomic2018}, while SOAP and SNAP were computationally intensive due to spherical harmonics and higher-order angular terms \cite{caroOptimizingManybodyAtomic2019,zuoPerformanceCostAssessment2020}. ACE provided a complete, scalable framework for atomic representation but introduced challenges with higher-order term computations \cite{dussonAtomicClusterExpansion2022}.

HDNNs demonstrated near-DFT accuracy in predicting energy-related properties for elemental crystals with simple lattice types. However, these models require individual training for specific chemical systems and struggle with binary, ternary, or quaternary systems featuring complex lattices, limiting their transferability. Additionally, all four descriptors focused on local atomic environments within a cutoff radius, lacking a global encoding of the entire systemic geometry and topology. Efforts to incorporate long-range interactions, such as electrostatics \cite{yaoTensorMol0ModelChemistry2018,unkePhysNetNeuralNetwork2019} and nonlocal charge-transfer \cite{farajiHighAccuracyTransferability2017,farajiSurfaceReconstructionsPremelting2019,beckeMulticenterNumericalIntegration1988,xieIncorporatingElectronicInformation2020}, has greatly improved HDNN accuracy but still primarily targeted DFT energies and partial charges. Despite these advancements, accurate prediction of electronic properties, such as Fermi energy and band gap, remains an area requiring further exploration.

GNNs, on the other hand, eliminate the need for extensive descriptor engineering by directly learning from structural data. This capability enables GNNs to model both the local atomic environment and the global structure of materials in a more natural and scalable manner, making them particularly useful for applications in materials discovery, such as predicting electronic properties, phase transitions, and mechanical properties \cite{schuttSchNetDeepLearning2018,chenGraphNetworksUniversal2019a,choudharyAtomisticLineGraph2021b,geigerE3nnEuclideanNeural2022,yanPeriodicGraphTransformers2022,schuttSchNetPackNeuralNetwork2023,gurunathanRapidPredictionPhonon2023}. One of the most influential models in this area is the Crystal Graph Convolutional Neural Network (CGCNN) \cite{xieCrystalGraphConvolutional2018}, introduced by Xie and Grossman in 2018. CGCNN represents materials as graphs to capture the structural and bonding information of crystalline solids. This representation, combined with convolutional layers that update atomic and bond features through a message passing mechanism \cite{gilmerNeuralMessagePassing2017}, makes CGCNN highly effective in predicting material properties from crystal structures. The success of CGCNN sparked widespread interest in the use of GNNs for material property prediction.

Despite the success, distance-based message-passing GNN architectures like CGCNN face two major limitations. One limitation is that they only encode pairwise distance information for edge features, neglecting many-body interactions that are essential for accurately modeling atomic environments \cite{parsaeifardManifoldsQuasiconstantSOAP2022}. Many-body interactions involve complex dependencies among multiple atoms, where the properties of an atom cannot be fully described by its interactions with just one or two neighbors. Another challenge is that current GNN-based machine learning models often require extensive human intervention and numerous \textit{in silico} experiments to select appropriate chemical information (such as atomic number, valence electrons, electronegativity, covalent radius) for node feature encoding. This manual feature selection limits the model performance and transferability, particularly when working with smaller datasets \cite{2020AcharyaFeaSele,dwivediBenchmarkingGraphNeural2023}.

To address the first limitation, researchers have enhanced GNN architectures by incorporating extra geometric information and improving how these models process atomic interactions. One key improvement involves integrating more complex geometric features beyond simple bond lengths, such as angular and directional information. Models like DimeNet \cite{gasteiger_dimenet_2020}, GemNet \cite{gasteiger_gemnet_2021}, ALIGNN \cite{choudharyAtomisticLineGraph2021}, and M3GNet \cite{chenUniversalGraphDeep2022a} leverage these geometric details to capture intricate spatial relationships between atoms, which is particularly important in anisotropic materials. Moreover, models like E3NN \cite{geigerE3nnEuclideanNeural2022} and NequIP \cite{batznerE3equivariantGraphNeural2022} incorporate vector information to ensure the representations remain invariant under rotations and translations, thereby enhancing the ability to model complex atomic systems.
In addition, attention mechanisms and global aggregation techniques have revolutionized how GNNs handle information flow. Attention layers, used in models like GATGNN \cite{louisGraphConvolutionalNeural2020} and Equiformer \cite{liao2023equiformer}, dynamically adjust the importance of different atomic interactions, allowing the network to emphasize the most chemically relevant bonds. These attention mechanisms enhance local expressiveness and improve the aggregation of global structural information. By combining these features with advanced readout functions---such as global attention and hierarchical pooling---models like MEGNet \cite{chenGraphNetworksUniversal2019} and GraphTrans \cite{wu2021graphtrans} ensure that critical information is preserved across the entire structure. Together, these innovations allow GNNs to efficiently prioritize and synthesize both local and global atomic interactions, improving the accuracy of material property predictions. Beyond these general developments, there have also been efforts to tailor GNNs specifically for crystalline materials. For example, GeoCGNN \cite{chengGeometricinformationenhancedCrystalGraph2021} and Matformer \cite{yanPeriodicGraphTransformers2022} explicitly encode crystal periodicity, while the Reciprocal Space Neural Network (RSNN) \cite{yuCapturingLongrangeInteraction2022} leverages reciprocal space information to capture long-range interactions in periodic systems. These developments underscore the growing complexity and sophistication of GNN architectures in materials science, enabling more accurate and generalizable property predictions.

In this work, we introduce EOSnet (Embedded Overlap Structures for Graph Neural Networks) to overcome the limitations in GNN models for materials science. EOSnet adopts a novel approach to incorporate many-body interactions through Gaussian Overlap Matrix (GOM) fingerprints \cite{sadeghiMetricsMeasuringDistances2013,zhuFingerprintBasedMetric2016} as atomic features within the GNN architecture. Unlike previous GNN models that embed the angular information into edge features explicitly, EOSnet uses GOMs to represent the overlap of atomic orbitals between neighboring atoms, providing a compact and efficient description of the many-body interactions within a material. Moreover, the fingerprint feature in EOSnet is designed to be generic and transferable, ensuring that two individual atoms will only share the same node feature if they have identical neighboring environments, even within the same atomic species. Another advantage of the GOM-based fingerprints is their rotational and translational invariance, ensuring a consistent description of atomic environments regardless of the structural orientation or position.  By embedding these fingerprints as node features, EOSnet achieves superior predictive performance across a range of material property tasks while maintaining computational efficiency.

In EOSnet, the crystal structure is represented as a graph, where each atom is a node, and the edges represent bonds between neighboring atoms. The model architecture, depicted in Figure 1, integrates GOM-based fingerprints within a graph convolutional neural network (GCNN) framework. These GOM fingerprints are derived from the eigenvalues of Gaussian overlap matrices, designed to capture many-body atomic interactions within a cutoff sphere centered around each atom in the crystal structure. 
To construct the GOM, Gaussian-type orbitals (GTOs) are centered on each atom within this sphere, and the overlap integrals are calculated between every pair of atoms within the cutoff radius. The width of each Gaussian function is determined by the covalent radius of the atom on which it is centered, incorporating key features of the covalent radius into the atomic representation. Additionally, a cutoff function ensures that the overlap integrals smoothly decay to zero at the boundary of the sphere, preventing discontinuities when atoms enter or leave the region.

The fingerprint for each atom is then computed by extracting the eigenvalues of its corresponding GOM. These eigenvalues encode essential information about atomic environments and serve as rotation and translation invariant features for the graph neural network. This design captures the strength of interactions between an atom and its neighbors, as well as the interactions among the neighboring atoms themselves, offering a comprehensive and efficient representation of many-body atomic interactions.  Since the Gaussian parameters used in the GTOs are fixed, the descriptor construction process is simplified, avoiding extra tuning for specific systems.
For moderate-sized systems, GOM-based fingerprints offers a balance between expressiveness and computational efficiency.
Unlike methods that rely on costly angular expansions, such as SOAP and bispectrum descriptors, GOMs produce a matrix (or eigenvalue) representation that can be directly fed into machine learning models without extensive preprocessing. This straightforward mathematical structure allows neural networks to capture both local and global structural properties effectively.
Figure 2 provides an example of the GOM fingerprints for a cubic \ch{SrTiO3} structure, illustrating how the overlap strength between atomic orbitals is encoded in the GOM. This figure highlights the many-body interactions and shows how the GOM captures the collective behavior of neighboring atoms.


The EOSnet architecture follows a standard message-passing framework, where the node features (GOM fingerprints) are updated based on the interactions between neighboring atoms. The workflow of EOSnet is illustrated in Figure 3.
In each convolutional layer, the node feature for atom $i$ is updated by aggregating information from its neighboring atoms $j$ through a message-passing mechanism. This process is defined as:

\begin{subequations}
    \begin{equation}
    \mathbf{n}_i^{(l+1)}=\mathbf{n}_i^{(l)}+\sum_{j, k} \sigma\left(\mathbf{W}_{\text {g }}^{(l)} \mathbf{m}_{(i, j)_k}^{(l)} + \mathbf{b}_{\text {g }}^{(l)} \right) \odot g\left(\mathbf{W}_{\text {m }}^{(l)} \mathbf{m}_{(i, j)_k}^{(l)} + \mathbf{b}_{\text {m }}^{(l)}\right)
    \tag{1}\label{eq:1}
    \end{equation}
    \begin{equation}
    \mathbf{m}_{(i, j)_k}^{(l)}=\mathbf{n}_i^{(l)} \oplus \mathbf{n}_j^{(l)} \oplus \mathbf{e}_{(i, j)_k}
    \tag{2}\label{eq:2}
    \end{equation}
\end{subequations}

Here, $\mathbf{m}_{(i, j)_k}^{(l)}$ is the message from atom $j$ to atom $i$ via the $k$-th edge, $\mathbf{n}_i^{(l)}$ denotes the node feature of atom $i$ at layer $l$, $\mathbf{e}_{(i, j)_k}$ represent the $k$-th edge feature between atom $i$ and $j$. And $\mathbf{W}_{\text {g }}^{(l)}$, $\mathbf{b}_{\text {g }}^{(l)}$ are the gate weight and bias at layer $l$, $\mathbf{W}_{\text {m }}^{(l)}$, $\mathbf{b}_{\text {m }}^{(l)}$ are the message weight and bias at layer $l$. Operator $\odot$ is the element-wise matrix multiplication, $\oplus$ is the vector concatenation, $\sigma$ and $g$ are independent nonlinear activation functions.

The convolutional layers, as depicted in Figure 1, process the GOM-embedded node features, allowing the network to retain information about the many-body interactions throughout the depth of the model. After each message-passing step, node features are updated using gated convolutional layers. These layers ensure that relevant information from the neighbors is retained while irrelevant information is filtered out. This process is repeated for several layers, allowing the model to capture both local and global atomic interactions.

At the end of the graph convolutional layers, the node features are aggregated using a global pooling operation to predict the target property. In EOSnet, we use an average pooling function, which computes the mean of all node features in the graph, followed by a fully connected layer that maps the aggregated feature vector to the target property:
\begin{equation}
    \text { Output }=\operatorname{FCN}\left(\frac{1}{N_{\text{at}}} \sum_{i=1}^{N_{\text{at}}} \mathbf{n}_i^{(l^*)}\right)
    \tag{3}\label{eq:3}
\end{equation}
where $N_{\text{at}}$ is the number of atoms in the crystal structure, and $\mathbf{n}_i^{(l^*)}$ is the final node feature after the last convolutional layer. The fully connected layer (FCN) maps the aggregated node features to the target property, providing the final prediction.  

In addition to the GOM fingerprints used for node features, EOSnet also utilizes bond distance expansions to define the edge features between atoms. These edge features are expanded using Gaussian filters to ensure smooth and continuous representation of interatomic distances. The Gaussian distance expansion of the $k$-th bond is expressed as:
\begin{equation}
    \tilde{D}(i,j,k) = \exp \left( -\frac{(\left\|\mathbf{r}_{i, j}\right\| - d_k)^2}{\sigma_k^2} \right) 
    \tag{4}\label{eq:4}
\end{equation}
where $\left\|\mathbf{r}_{i, j}\right\|$ is the distance between atoms $i$ and $j$, $d_k$ is the center of the Gaussian filter and $\sigma_k$ the scale
parameter controlling the Gaussian width. These expanded bond distances provide additional geometric information for the convolutional layers, enabling the model to effectively learn from both node and edge features. 
The inclusion of these features is crucial for capturing short-range interactions, especially in systems where the geometric arrangement of atoms significantly impacts material properties. The workflow for this process is also detailed in Figure 3.

EOSnet is trained on a variety of material datasets, including both large ($>$ 20,000) and small ($<$ 2,000) datasets for material property prediction. The model is trained using a supervised learning approach with an 80\%-10\%-10\% train-validation-test split. Mean absolute error (MAE) is used as the loss function for regression tasks, such as predicting formation energy, total energy, and bandgap. For classification tasks, such as distinguishing between metals and non-metals, cross-entropy loss is used. A dynamic class weight scheme is implemented to eliminate the effect of class imbalance in binary classification problems. The model is optimized using the Adam optimizer with a learning rate scheduler to dynamically adjust the learning rate during training. Gradient clip technique is used to prevent over-fitting during training process.

\begin{figure*}
    \centering
    \includegraphics[width=\textwidth]{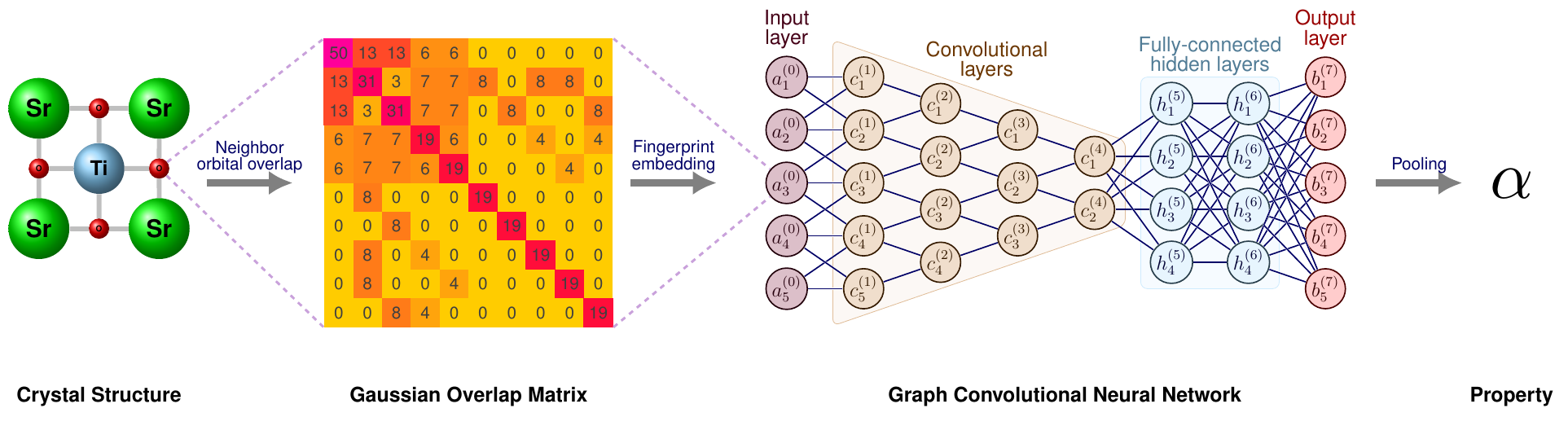}
    \caption{Schematic architecture of EOSnet, illustrating the workflow from the crystal structure to property prediction. The left panel shows a crystal structure of SrTiO$_3$, used to generate the Gaussian Overlap Matrix (GOM) that captures neighbor orbital overlaps. The GOM eigenvalues are embedded as input node features for the graph convolutional neural network, depicted in the middle. The enclosed cyan box highlights the convolution and message-passing processes. The right panel shows fully connected layers and a pooling operation, ultimately predicting the material property, denoted by $\alpha$.}
    \label{fig:FpGNN_scheme}
\end{figure*}

\begin{figure*}
    \centering
    \begin{subfigure}[b]{0.55\textwidth}
        \centering
        \includegraphics[width=\textwidth]{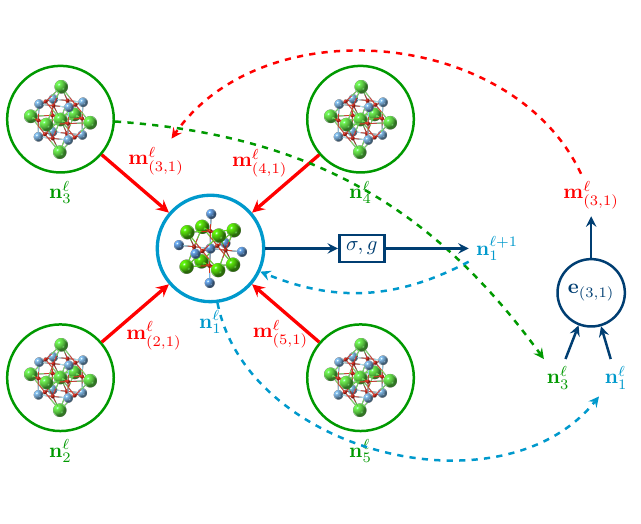}
        \caption*{}
        \label{fig:EOSnet_MP_GCNN_Scheme}
    \end{subfigure}
    \hfill
    \begin{subfigure}[b]{0.40\textwidth}
        \centering
        \raisebox{25pt}{\includegraphics[width=\textwidth]{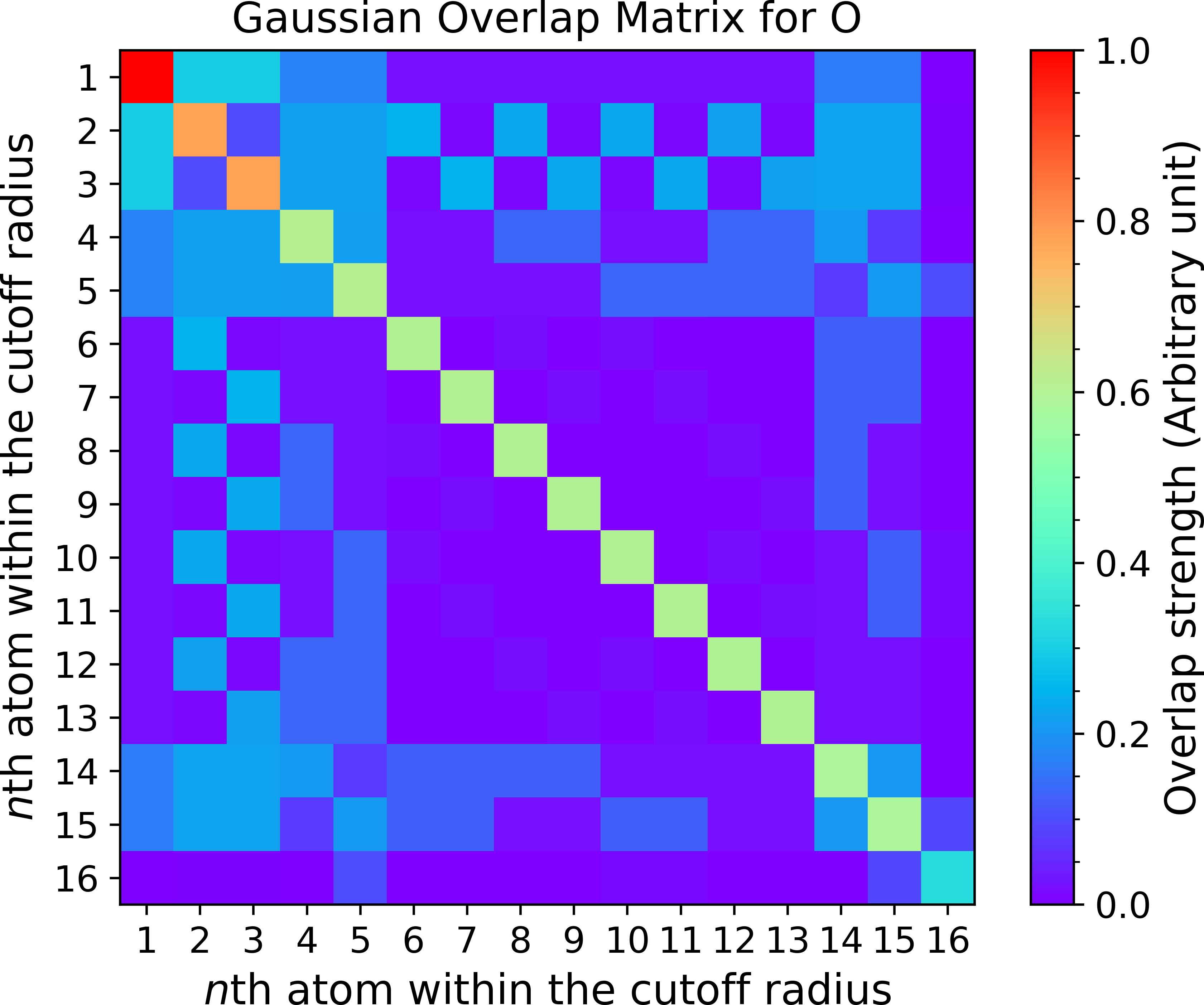}}
        \caption*{}
        \label{fig:GOM_heatmap}
    \end{subfigure}
    \vspace{-25pt}
    \caption{Illustration of many-body interaction and GOM. Left: Demonstration of the feature aggregation and message-passing scheme for an SrTiO$_3$ crystal using GOM fingerprints. Green, blue, and red spheres represent Sr, Ti, and O atoms, respectively. For simplicity, only four Sr atoms are considered as nearest neighbors to the Ti atom. Right: Normalized Gaussian Overlap Matrix for an O atom, with atoms within the cutoff radius sorted by their distance from the central atom. The (1,1) element is always 1.0, indicating self-interaction. For demonstration purposes, we use an 8.0 {\AA} cutoff radius, considering only s-orbitals and the first 16 neighboring atoms.}
    \label{fig:Fp_details}
\end{figure*}

\begin{figure*}
    \centering
    \includegraphics[width=0.5\textwidth]{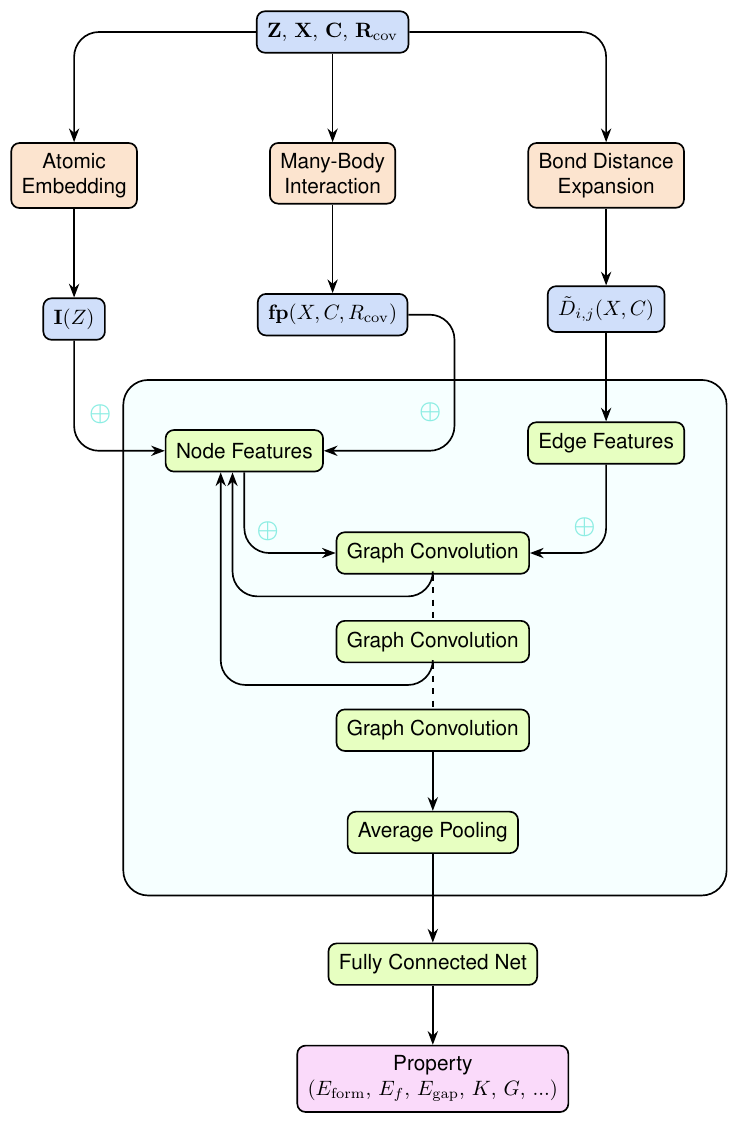}

    \caption{Workflow of EOSnet, which uses fractional coordinates (\( \mathbf{X} \in \mathbb{F}^{N \times 3} \), where \( \mathbb{F} = \{ f \in \mathbb{R} : 0 \leq f < 1 \} \)), lattice vectors (\( \mathbf{C} = [\mathbf{c}_1, \mathbf{c}_2, \mathbf{c}_3] \in \mathbb{R}^{3 \times 3} \)), atomic numbers (\( \mathbf{Z} \)), and covalent radii (\( \mathbf{R}_{\text{cov}} \)) as inputs. The \( \tilde{D}_{i,j} \) denotes bond distance expansion using Gaussian filters, defined in Equation \ref{eq:4}. The process incorporates atomic embedding, many-body interaction, and bond distance expansion to generate node and edge features, which are processed through graph convolution layers and pooling to predict material properties. }
    \label{fig:FpGNN_flowchart}
\end{figure*}

The performance of the proposed EOSnet model was evaluated across various material property prediction tasks, including the prediction of total energy, formation energy, Fermi energy, band gap, bulk modulus, shear modulus, and a metal/non-metal classification task. To benchmark the effectiveness of EOSnet, comparisons were made with several models, including CGCNN \cite{xieCrystalGraphConvolutional2018}, MEGNet\cite{chenGraphNetworksUniversal2019}, and M3GNet\cite{chenUniversalGraphDeep2022a}. The summarized results are presented in Table 1. It is important to note that although all models listed in Table 1 were trained on datasets from the Materials Project \cite{jainCommentaryMaterialsProject2013}, the sizes of the datasets and the evaluation metrics may differ. The primary focus of this comparison is not to claim that EOSnet is universally superior but to highlight its performance across various tasks. Specifically, we emphasize the advancements made by EOSnet in capturing many-body interactions to enhance the prediction accuracy of material properties. Studies show that the MAE of ML models generally decreases with the increase in the size of the training dataset\cite{chenGraphNetworksUniversal2019, chenUniversalGraphDeep2022a,dunnBenchmarkingMaterialsProperty2020}. Therefore, the performance of EOSnet is expected to improve further with larger training datasets in the future studies.  

Figure 4a and 4b provide a detailed analysis of the EOSnet model performance in predicting formation energy and electronic band gap.  The color gradient in these figures represents the density of data points, with brighter colors indicating regions of higher density. The tight clustering of predicted values around the diagonal line indicates that EOSnet effectively models the distribution of formation energy and band gap values, demonstrating robustness and reliability. 
An example demonstrating the strengths of EOSnet is its band gap prediction, achieving a low MAE of 0.163 eV.  This performance is impressive comparing to other models, despite differences in the size of training datasets used. Band gap prediction is particularly challenging due to its sensitivity to many-body interactions, as well as electronic correlations within the material. Models that primarily rely on bond distances and basic geometric features often struggle to capture the nuanced effects that influence the electronic structure.
The use of GOM-based fingerprints in EOSnet enables a more comprehensive representation of many-body interactions. This method captures not only the local interactions between an atom and its immediate neighbors but also the interactions among neighboring atoms themselves. 
For example, in semiconducting materials, the band gap, defined by the energy separation between valence and conduction bands, emerges from complex interplays between atomic arrangements, chemical bonding, and the electronic environment.  While conventional GNNs often struggle to capture subtle electronic interactions and intricate orbital overlaps that determine band gaps, EOSnet effectively encodes these crucial quantum mechanical features.

The enhanced performance extends beyond band gap prediction to other properties sensitive to many-body interactions, such as bulk and shear moduli (Table 1). This broad success demonstrates the effectiveness of EOSnet approach to capturing complex atomic interactions in materials. The model also excels in classification tasks, achieving 97.7\% accuracy in metal/non-metal classification and 94.5\% accuracy in predicting the dynamic stability of guest-atom substituted type-VII boron-carbide clathrates (Figure 4c and 4d).
Particularly noteworthy is the model strong performance in stability classification despite limited training data, highlighting how the non-human-engineered atomic features can extract meaningful patterns even from small datasets.

In summary, we have introduced EOSnet (Embedded Overlap Structures for Graph Neural Networks), a new approach that enhances the predictive capabilities of graph neural networks in materials science by efficiently incorporating many-body interactions through GOM-based fingerprints. EOSnet provides a rotationally invariant and computationally efficient method to represent the full spectrum of atomic interactions without the need for explicit higher-order terms. Our extensive evaluations across various material property prediction tasks demonstrate the performance of EOSnet, particularly in predicting properties that are sensitive to many-body interactions such as electronic band gap and elasticity. Notably, EOSnet achieved a MAE of 0.163 eV in band gap prediction, surpassing the original CGCNN model and even better than M3GNet with the number of training data increasing over a order of magnitude. Same performance boost has been observed in the case of bulk modulus and shear modulus. This indicate embedding GOM fingerprints into node features can indeed helps the GNN model understand these properties better with the essential information of atomic environments, including the collective behavior of neighboring atoms and the strength of their orbital overlaps. This improved performance is especially important for applications in material discovery, making it a valuable tool for the design of materials with optimized electronic, mechanical, and thermal properties.
While EOSnet shows performance improvements, further work could include expanding the model to handle more diverse datasets and incorporating additional geometric or 
attention mechanisms to capture long-range interactions more effectively. Additionally, exploring its applicability to other domains, such as catalysis or battery materials, could open new avenues for material discovery.

\begin{figure*}
    \centering
    \includegraphics[width=\textwidth]{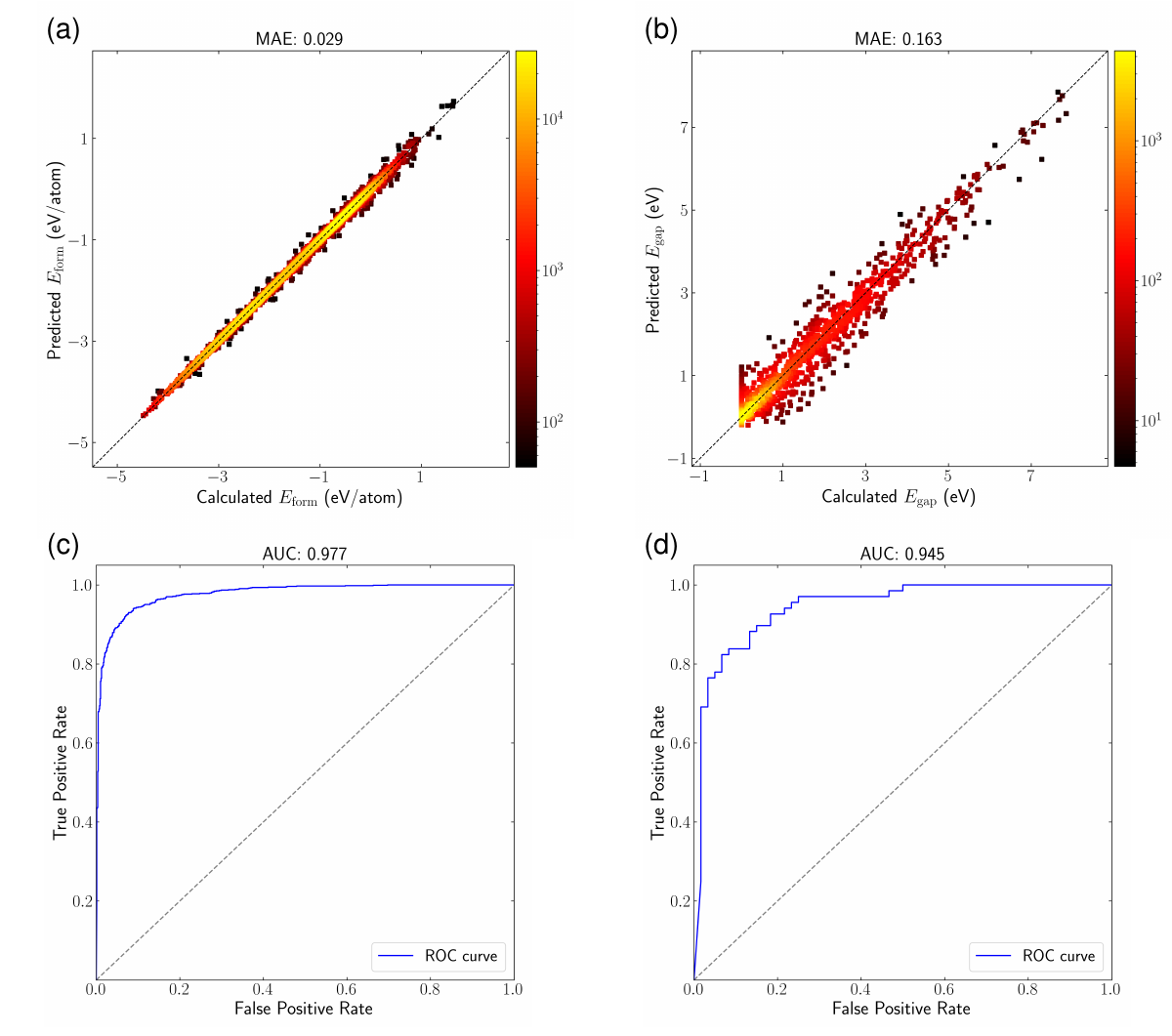}
    \caption{Parity plots and ROC curves for evaluating EOSnet performance. (a) Parity plot for the prediction of electronic band gap ($E_g$) using 19,393 data points from the Materials Project, showing an MAE of 0.163 eV. (b) Parity plot for formation energy predictions on 50,654 data points from the Materials Project, with an MAE of 0.029 eV. (c) ROC curve for metal/non-metal classification, achieving an AUC of 0.977. (d) ROC curve for dynamically stable/unstable classification on 1,335 guest-atom-substituted type-VII boron-carbide clathrates (MB$_{6-x}$C$_x$, $x$ from 1 to 5), achieving an AUC of 0.945. The clathrate stability is determined using phonon frequency from VASP\cite{kresseEfficientIterativeSchemes1996,kresseUltrasoftPseudopotentialsProjector1999} calculations.}
    \label{fig:Data_plot}
\end{figure*}

\begin{table}[!ht]
\centering 
\begin{minipage}{\textwidth}
\centering
\caption{Comparison of the MAEs in Total Energy ($E_{\text{tot}}$), Formation Energy ($E_{\text{form}}$), Fermi Energy ($E_{f}$), Band Gap ($E_g$), Bulk Modulus ($K_{\text{VRH}}$), Shear Modulus ($G_{\text{VRH}}$) and Metal/Nonmetal Classification between EOSnet and previous works. The number of structures in the training data is in parentheses.}
\begin{tabular}{|c|c|c|c|c|c|}
\hline
\textbf{} & \textbf{EOSnet} \footnote{For our model, the numbers in parentheses indicate the total numbers of structures before train-validation-test split. 19,364 $E_{\text{tot}}$ data are from ASE \cite{Hjorth-Larsen-2017} cubic perovskite dataset \cite{castelliComputationalScreeningPerovskite2012,castelliNewCubicPerovskites2012,castelliCalculatedPourbaixDiagrams2014}. 27,293 $E_{f}$ data and 19,393 band gap data are from Materials Project 2024 dataset. 5000 Voigt-Reuss-Hill (VRH) average bulk moduli, shear moduli and 50654 $E_{\text{form}}$ data are from Materials Project 2018 dataset.} & \textbf{CGCNN} \cite{xieCrystalGraphConvolutional2018} & \textbf{M3GNet} \cite{chenUniversalGraphDeep2022a} & \textbf{MEGNet} \cite{chenGraphNetworksUniversal2019} \\ \hline
\multirow{2}{*}{$E_{\text{tot}}$ (eV$\cdot$atom$^{-1}$)} & 0.034 & 0.072 & 0.035 & \multirow{2}{*}{-} \\ 
                    & (19,364) & (28,046) & (132,752) &  \\ \hline
\multirow{2}{*}{$E_{\text{form}}$ (eV$\cdot$atom$^{-1}$)} & 0.029 & 0.039 & 0.020 & 0.028 \\ 
                    & (50,654) & (28,046) & (132,752) & (60,000) \\ \hline
\multirow{2}{*}{$E_{f}$ (eV)} & 0.295 & 0.363 & \multirow{2}{*}{-} & \multirow{2}{*}{-} \\ 
                    & (27,293) & (28,046) & &  \\ \hline
\multirow{2}{*}{$E_g$ (eV)} & 0.163 & 0.388 & 0.183 & 0.33 \\ 
                    & (19,393) & (16,485) & (106,113) & (36,720) \\ \hline
\multirow{2}{*}{$K_{\text{VRH}}$ (log$_{10}$ (GPa))} & 0.034 & 0.054 & 0.058 & 0.050 \\ 
                    & (5,000) & (2,041) & (10,987) & (4,664) \\ \hline
\multirow{2}{*}{$G_{\text{VRH}}$ (log$_{10}$ (GPa))} & 0.072 & 0.087 & 0.086 & 0.079 \\ 
                    & (5,000) & (2,041) & (10,987) & (4,664) \\ \hline
\multirow{2}{*}{Metal/Nonmetal Classifier} & 97.7\% & 95.0\% & 95.8\% & 90.6\% \\ 
                    & (19,393) & (16,458) & (106,113) & (55,391) \\ \hline
\end{tabular}
\end{minipage}
\end{table}

\begin{acknowledgement}

This work was supported by the National Science Foundation, Division of Materials Research (NSF-DMR) under Grant No. 2226700, and startup funds of the office of the Dean of SASN of Rutgers University-Newark. The authors acknowledge the Office of Advanced Research Computing (OARC) at Rutgers, The State University of New Jersey, for providing access to the Amarel cluster and associated research computing resources that have contributed to the results reported here.

\end{acknowledgement}





\bibliography{mainbib}

\end{document}